\documentclass{amsart}
\usepackage{amsmath}
\usepackage{amsfonts}
\usepackage{amssymb}
\usepackage{amsthm}
\usepackage{amsopn}
\usepackage{amscd}
\usepackage{amsxtra}
\usepackage[foot]{amsaddr}
\usepackage{hyperref}

\usepackage{tabularx,ragged2e,booktabs,caption}

\newtheorem{theorem}{Theorem}[section]

\theoremstyle{definition}
\newtheorem{definition}[theorem]{Definition}

\theoremstyle{remark}

\newcommand{\RR}{\mathbb{R}}
\newcommand{\NN}{\mathbb{N}}

\newcommand{\thb}{\mathbf{\theta}}
\newcommand{\thtrue}{ {\mathbf{\theta}^*} }
\newcommand{\thhat}{\widehat{\thb}}

\newcommand{\gbtwo}{\textrm{GB2}}

\newcommand{\uomten}{UOM1}

\author{Paul Larsen}

\email{paul.larsen@maths.oxfordalumni.org}
\title[Operational risk models and asymptotic normality]{Operational risk models and asymptotic normality for small sample-sizes}
\date{\today}
\numberwithin{equation}{section}

\usepackage{Sweave}
\begin{document}

\maketitle
\tableofcontents

\begin{abstract}
Operational risk models commonly employ maximum likelihood estimation (MLE) to fit loss data to heavy-tailed distributions. Yet several desirable properties of MLE (e.g. asymptotic normality) are generally valid only for large sample-sizes, a situation rarely encountered in operational risk. In this paper, we study how asymptotic normality does--or does not--hold for common severity distributions in operational risk models. We then apply these results to evaluate errors caused by failure of asymptotic normality in constructing confidence intervals around the MLE fitted parameters. 
\end{abstract}


\section{Introduction}
Maximum likelihood estimation (MLE) is a--if not the--standard method for fitting parametric distributions in operational risk models \cite{thorny}. Its widespread use is due in large part to properties that hold as the sample-size of loss data goes to infinity, namely, that MLE is a consistent, asymptotically normal, and asymptotically efficient estimator. We focus here on the second property, asymptotic normality of MLE, and how this property relates operational value-at-risk (OpVaR) models. Informally, asymptotic normality of MLE means that the estimated parameters will be normally distributed about the true parameters with variance going to zero as the sample-size tends to infinity (greater detail will be given in Section \ref{sec:basics}).

The assumption of sample-sizes approaching infinity, however, is hard to justify in operational risk, where capital estimates are driven by large and rare loss events, thus making the asymptotic nature of asymptotic normality ground for concern in OpVaR models. The situation is summarized in \cite{embrechts1997modelling}, p. 318: ``[A]lthough we have reliable numerical procedures for finding the MLE $\ldots$, we are less certain about its properties, especially in the small sample case.''

The challenge of small sample-sizes for MLE estimation is exacerbated by pressure to develop stand-alone OpVaR models. Rather than calculating a single OpVaR for all of a given bank's legal entities, and then sub-allocating this capital figure to the legal entities, some local regulators are asking for OpVaR models that calculate on the level of a given legal entity (or cluster of entities in a country). Hence the problem of fitting a heavy-tailed distribution to a relatively small number of high losses across a bank has been made more severe by effectively carving the bank--and its loss data--into even smaller pieces.

In this first of two papers, we study how--if at all--asymptotic normality holds for heavy-tailed distributions common to OR modeling, and what this means for OR modeling. Specifically we 
\begin{enumerate}
\item  \label{enum:plots} perform graphical and numerical tests of normality (Sections \ref{sec:graphical} and \ref{sec:numerical}), 
\item \label{enum:conf} assess the normal approximation for parameter fitting confidence intervals (Section \ref{sec:anCIs}).
\end{enumerate}
The second point refers to using asymptotic normality to determine confidence intervals for parameter estimates. These confidence intervals are then used for goodness-of-fit tests when fitting tail distributions. Of course, these estimates are only reliable insofar as the parameter error is distributed as predicted by asymptotic normality.  In a second paper \cite{mleor_var}, we investigate how MLE error translates into stability of OpVaR models (or a lack thereof).

While this paper focuses on operational risk, the problem of estimating errors resulting from MLE fitting of heavy-tailed distributions is much more general. The heavy-tailed distributions we consider (Pareto, Weibull, lognormal, log-logistic and generalized Beta of the second kind) arise in many other contexts besides operational risk, such as data networks, market models and insurance \cite{resnick2007heavy}. Indeed, the type of OpVaR model considered here is commonly called the ``actuarial approach,'' and is characterized by modeling frequencies and severities of losses separately. 

Relating theoretical concerns to practice requires representative loss data. We use loss data from the recent ORX Association OpVaR stability study, in which participants were given loss data sets for 12 units of measure (UOM) consisting of anonymized loss data from member banks. Four UOMs were selected for this paper that spanned a range of loss profiles. Due to space considerations, analyses are given here for only one UOM (\uomten), with full results for this UOM and three others given in a separate appendix, \cite{mleorappendix}.

\noindent \textbf{Main findings} \newline
We now summarize our main findings.
\begin{itemize}
\item In Section \ref{sec:basics}, we present the theory of asymptotic normality for the Pareto, Weibull, lognormal, log-logistic and GB2 distributions, and show how the typical assumptions of asymptotic normality are difficult, if not impossible, to verify for these severity distributuions.
\item In Section \ref{sec:graphical} we test asymptotic normality graphically for small sample sizes via simulated MLE parameters (parametric bootstrapping). These tests show poor results for the Weibull and GB2 distributions, while the Pareto, lognormal and log-loglogistic distributions seem to display asymptotic normality, even with small samples sizes. 
\item Numerical tests of asymptotic normality are studied in Section \ref{sec:numerical}, where the resulting p-values indicate the lognormal distribution is only one for which the normality assumption is not completely rejected.
\item Despite the departures from normality that can be seen graphically and numerically, asymptotic normality gives a remarkably good approximation to MLE confidence intervals for the Pareto, lognormal and log-logistic distributions (Section \ref{sec:anCIs}). The performance for the Weibull and GB2 distributions is quite poor.
\end{itemize}
\noindent The use of asymptotic normality to estimate MLE error is common in practice, and is also used in \cite{frachot2004loss, cope2009challenges} to estimate parameter confidence intervals. Possible shortcomings of this approximation are mentioned in \cite{fabio}.


To conclude this section, we mention a few technical points. For the probability distributions considered below, different sources give different names for parameters, and sometimes the distribution functions themselves vary from source to source. We give precise definitions in the next section, and generally follow parameter naming conventions as in the \texttt{R} packages below that were used for our analyses.

The statistical analyses, graphics and typesetting were all done via the statistical software \texttt{R} \cite{R}. The \texttt{R} packages beyond the ones in the base set-up are 
\texttt{fitdistrplus} \cite{fdp},
\texttt{GB2} \cite{gb2},
\texttt{ggplot2} \cite{ggp2},
\texttt{neldermead} \cite{nm},
\texttt{VGAM} \cite{vgam}, 
\texttt{MVN} \cite{Korkmaz:2014fk}, and
\texttt{Sweave} \cite{lmucs-papers:Leisch:2002b, lmucs-papers:Leisch:2003b}.

\section{OpVaR, asymptotic normality of MLE and heavy-tailed distributions}
\label{sec:basics}

 \subsection{Asymptotic normality of MLE}
 \label{sec:an}
 Informally, asymptotic normality of MLE says that the distribution of fitted parameters to data will be normally distributed, centered about the true parameters, with a prescribed covariance matrix that depends on the sample-size. Let $X = (x_1, \ldots, x_n)$ be data from an underlying distribution with probability density function (PDF) $f(x|\thtrue)$, where $\thtrue$ are the true parameters. Then to test if asymptotic normality holds for this distribution, we need to test if, as $n$ increases, MLE yields fitted parameters $\thhat$ that are normally distributed. This property can be tested with parametric bootstrapping, that is, for a fixed $n$, we sample $n$ data points from the true distribution $f(x|\thtrue)$, and apply MLE to get $\thhat_{1,n}$. We repeat this sample/fit procedure $m$ times to obtain fitted parameters $(\thhat_{1,n}, \ldots,\thhat_{m,n})$, i.e. we have generated statistics for MLE fitted parameters for a sample size of $n$.

Before giving a precise statement of asymptotic normality for MLE, we first chronicle the regularity assumptions required for its proof. Define the \emph{log-likelihood} function of the distribution as
$\ell(x| \theta) = \log f(x|\theta)$. The notation $E_\thtrue\left[ g(x|\theta) \right]$ for a function $g(x| \theta)$ means
\begin{equation*}
E_\thtrue \left[ g(x|\theta) \right] = \int g(u|\thtrue) f(u | \thtrue) du.
\end{equation*}

Then the usual regularity conditions are \cite{cox1979theoretical, greene2011econometric}
\begin{enumerate}
\item \label{reg:1} The first three partial derivatives of $\ell(x|\theta)$ with respect to $\theta$ are continuous and finite for almost all $x$ and for all $\theta$ in a neighborhood of $\thtrue$.
\item \label{reg:2} For all $\theta_j$, and $i=1,2$, 
\begin{equation*}
E_\thtrue \left[\frac{\partial_i \ell(x|\theta)}{\partial (\theta_j)^i} \right] < \infty.
\end{equation*}
\item \label{reg:3} There exists a distribution function $M(x)$ such that $\| \frac{\partial_3 \ell(x|\theta)}{\partial \theta_k^3} \| < M(x)$ for all $\theta$ in a neighborhood of $\thtrue$, and $E_\thtrue\left[ M(x))\right] < \infty$. 
\end{enumerate}

The \emph{Fisher information matrix} for a $k$-parameter distribution is the $k\times k$ matrix whose $(i,j)$ entry is
\begin{equation*}
\left( I(\thtrue) \right)_{i,j} = E_\thtrue \left[ \left(\frac{\partial}{\partial \theta_i} \ell(x| \theta) \right) \left(\frac{\partial}{\partial \theta_j} \ell(x| \theta) \right) \right]
\end{equation*}

Given the regularity conditions above, the Fisher information matrix admits the following simpler description:
\begin{equation}
\left( I(\thtrue) \right)_{i,j} = -E_\thtrue \left[ \frac{\partial^2}{\partial \theta_i \partial \theta_j} \ell(x| \theta) \right]
\end{equation}

A further requirement for asymptotic normality is \emph{identifiability} (this requirement is usually invoked when discussing the \emph{consistency} of MLE, which states that the estimated parameters converge to the true parameters in probability as the sample-size goes to infinity). Informally, a parametric distribution family is identifiable if the parameter $\theta$ uniquely determines the distribution (i.e. no two different parameter values yield the same distribution). More precisely, a distribution family is \emph{identifiable} if for any $\theta_1 \neq \theta_2$, there exists $X = (x_1, \ldots, x_n)$ for some $n$ such that $f(X| \theta_1) \neq f(X| \theta_2)$.

Proving identifiability can be challenging, but if the moments of $f(x| \theta)$ have a nice form, one strategy is as follows: Suppose for a contradiction that $\theta \neq \theta'$, and $X\sim f(x|\theta)$, $X' \sim f(x|\theta')$. If further there exists a $k \in \NN$ such that $E[X^k] \neq E[X'^k]$, then it follows that there exists a subset of the parameters space $U$ of non-zero measure such that $f(x| \theta) \neq f(x | \theta')$ for all $x \in U$, and hence the distribution family is identifiable.

A further requirement is that the Fisher information matrix be non-singular in a neighborhood of $\thtrue$. The study of when this condition fails has led to recent interaction between statistical learning theory and geometry, since the Fisher information matrix can be interpreted as a metric on the parameter space. Work of Sumio Watanabe and others develops a theory reconstructing many of the desirable properties of MLE in the case of singular Fisher information matrices by resolution of singularities from algebraic geometry \cite{watanabe2009algebraic, watanabe2013widely}.

The final requirement for asymptotic normality to hold is that the model has to be correctly specified: if MLE is applied for one parametric distribution to fit data coming from a different distribution, then of course results appealing to the ``true'' parameters will be suspect. 

\begin{theorem}[Asymptotic normality of MLE]
 \label{thm:an}
 Under the conditions above, the MLE $\thhat$ is asymptotically normal:
 \begin{equation*}
 \sqrt{n}(\thhat - \thtrue) \stackrel{d}{\to} N(0, I(\thtrue)^{-1}),
 \end{equation*}
 where convergence is in distribution.
\end{theorem}
\noindent A proof of the theorem can be found in \cite{wald1943tests}; sketch proofs are more abundant (see e.g. \cite{cox1979theoretical}).
 
Our main interest is in the asymptotic nature of this result. For an example of what can go wrong for finite sample-sizes, consider data $(x_1, \ldots, x_n)$ sampled independently from the normal distribution $N(\mu, \sigma)$. Then MLE produces the estimator for the variance $\hat{\sigma}^2 = 1/n \sum_i^n(x_i - \hat{\mu})^2$, which is biased for finite $n$.
 
This theorem gives a natural interpretation of the Fisher information matrix, which informally encodes how much information about the distribution is contained in each of the parameter directions. For simplicity, assume that the Fisher information is diagonal. Then large entries in the Fisher information matrix (high levels of information) correspond in Theorem \ref{thm:an} to small variations for MLE parameter estimation. In fact, a standard method to estimate MLE variance in numerical solvers is to calculate the Fisher information matrix at the optimal parameters, and invert it as in Theorem \ref{thm:an}. As a corollary, such variance estimates are in general only valid insofar as Theorem \ref{thm:an} applies, in particular, under the assumption of large sample-sizes. 
 
The higher-order regularity conditions above can be challenging to check in practice, and lack an obvious statistical interpretation. It can happen that the conditions of Theorem \ref{thm:an} are not satisfied, yet asymptotic normality still holds \cite{lecam1970, smith1985maximum}. Moreover, if the Fisher information matrix is singular (and not identically of determinant 0), then the set of parameters for which it is singular is of co-dimension at least one in the space of parameters (this is the solution set of $\det I(\theta) = 0$). Hence for almost all parameters $\theta$, the Fisher information matrix will be non-singular. 
Particular care is thus warranted when applying MLE to the generalized Beta distribution of the second kind, described in the next section, since the assumptions about the Fisher information matrix are difficult to verify.

A further challenge in bridging theory and practice is that for all but a few distributions, the algorithms used to determine the optimal parameters are numerical, and may produce only a local maximum of the log-likelihood function. As we describe the severity distributions under consideration below, we will thus also sketch the algorithms used in MLE and their potential shortcomings.

\subsection{Heavy-tailed distributions}
\label{sec:heavyTailed}
The severity distributions used in OpVaR models are generally \emph{heavy-tailed}. We will take heavy-tailed to mean \emph{subexponential} (definition below), but several other definitions exist in the literature. For the convenience of the reader, we also sketch other common definitions and, where possible, relate them to one another.

\begin{definition}
Let $F$ be a cumulative distribution function with support in $(0, \infty)$. Then $F$ is \emph{subexponential} if, for all $n \geq 2$,
\begin{equation*}
\lim_{x \to \infty} \frac{\overline{F}^{n*}(x)}{\overline{F}(x)} = n,
\end{equation*}
where $\overline{F}(x) = 1 - F(x)$ is the tail, or survival function, and the numerator in the limit is the $n$-fold convolution of $\overline{F}(x)$.
\end{definition}
\noindent Subexponential distributions exhibit one of the general properties expected of heavy-tailed distributions on the level of aggregate losses, namely that the tail of the maximum determines the tail of the sum. All of the distributions considered here are subexponential (for the Weibull distribution, this holds when the shape parameter is less than one; see Section \ref{sec:weibull} below). 

Subexponentiality implies another property that is sometimes taken as the definition of heavy-tailed, namely that the tail decays more slowly than any exponential function. With the notation as above, the precise formulation is that for all $\epsilon > 0$,
\begin{equation*}
\lim_{x \to \infty} e^{\epsilon x} \bar{F}(x) = \infty.
\end{equation*}
See \cite{embrechts1997modelling}, Lemma 1.3.5 (b) for the proof that a subexponential distribution function satisfies the above limit. 

An important subclass of subexponential distributions consists of \emph{regularly varying} functions:
\begin{definition}
A positive, Lebesgue measurable function $f$ on $(0, \infty)$ is \emph{regularly varying} at $\infty$ with index $\alpha \in \RR$ if 
\begin{equation*}
\lim_{x \to \infty} \frac{f(tx)}{f(x)} = t^\alpha
\end{equation*}
for all $t > 0$.
\end{definition}
\noindent Note that the lognormal and Weibull distributions are \emph{not} regularly varying.

For a regularly varying $F$ with tail index $\alpha > 0$, all moments of the associated random variable higher than $\alpha$ will be unbounded; see \cite{embrechts1997modelling} Proposition A.3.8 (d). Hence regular variation implies one final characterization of heavy tails. Some sources differentiate between heavy and ``light''-tailed distributions based on the existence of finite moments \cite{de2007implications}. Under this classification, the lognormal and Weibull distributions are light-tailed since all moments are finite, while the Pareto, log-logistic and GB2 distributions all have infinite moments, and are considered in this usage to be heavy-tailed.

One subclass of severity distributions we do not consider below arises from \emph{Extreme Value Theory}, such as the the Generalized Extreme Value (GEV) distribution.\footnote{We implicitly treat the Generelized Pareto Distribution (GPD), since for heavy-tailed distributions the GPD reduces to the Pareto distribution.} These distributions are generally not calibrated via MLE, but rather with methods from Extreme Value Theory, such as Peaks-Over-Threshold (see \cite{embrechts1997modelling}). Furthermore, consensus seems to have turned against EVT in operational risk (see e.g. \cite{mignola2005tests} for stability concerns with EVT). We thus limit our distributions to heavy-tailed distributions for which MLE is a prominent fitting method. Besides Extreme Value Theory distributions, we also pass over the g-and-h distribution, despite recent attention in the literature, since there is no
closed-form for its PDF, and is most naturally fitted to data by
quantile-matching \cite{taleTails} (MLE fitting methods do exist, however \cite{gnhMLE}).

The supports of the distribution families considered here may vary, which poses a problem when fitting loss data, especially in the case of a spliced distributions we study. For the Pareto distribution, one of the parameters defines the support, which contradicts an assumption required for the proof of asymptotic normality of MLE. Hence we set the parameter to the splice location, $T$, thus making the Pareto distribution a one-parameter family.

For the lognormal, log-logistic and Generalized Beta distribution of the second kind, the support is the positive real numbers. For fitting a spliced distribution, there are two standard approaches: replace the distribution with either the shifted or the truncated version to ensure that the support is contained in the tail region of the spliced distribution. Truncated distributions can pose difficulties for MLE fitting \cite{opdyke2012estimating}, plus explicit expressions for the resulting Fisher information matrices are more complicated, though in principle possible to obtain \cite{escobar1998fisher}. For these reasons, we consider exclusively shifted versions of these distributions.

\subsubsection{Pareto Distribution}

As mentioned above, the Pareto distribution is typically defined as a 2-parameter family, but since its support depends on one parameter (typically called the \emph{scale} parameter, $T$), we fix this as the threshold of our spliced severity distribution (e.g. $T = 100000$), and consider the Pareto distribution as depending on one parameter, the \emph{shape}, $\alpha$, resulting in PDF

\begin{equation*}
f(x| \alpha) = \frac{\alpha T^\alpha}{x^{\alpha + 1}},
\end{equation*}
where $x \geq T$, and is 0 otherwise. For $X \sim Pareto(\alpha)$, note that the first moment of $X$ is bounded if and only if $\alpha > 1$, and the variance is bounded only for $\alpha > 2$.

It is easy to show that the Pareto distribution is identifiable for all values of $\alpha$, and the Fisher information matrix is the scalar $I(\theta) = I(\alpha) = 1/\alpha^2$, which is indeed a positive-definite matrix (of size $1\times 1$). The unique solution to the likelihood equation $\nabla \ell = 0$ is 
\begin{equation*}
\widehat{\alpha} = \frac{n}{\sum_{1}^n \log(x_i/T)},
\end{equation*}
hence no numerical solver is required to perform MLE for the Pareto distribution.


\subsubsection{Weibull distribution}
\label{sec:weibull}
The Weibull distribution is a generalization of the exponential
distribution. For shape and scale parameters $a,b > 0$, the PDF is
\begin{equation*}
f(x|a,b) = \left(a/b\right) \left(x / b \right)^{a-1} \exp\left(-
  \left(\frac{x}{b} \right)^a\right)
\end{equation*}

In \cite{wei2007quantification}, the three-parameter Weibull distribution is considered,
with an extra location parameter $u$ that determines the support. As
noted above with reference to the Pareto distribution, a key
assumption of MLE is that the support is independent of the parameters
to be estimated. We thus consider the shifted distribution
for Weibull, which is equivalent to setting the location parameter to $u=T$.

The Fisher information matrix for the Weibull distribution is
\begin{equation}
I(\theta) = I(a, b)=
\begin{pmatrix}
\frac{1}{a^2}\left(\psi'(1) + \psi^2(2)\right) & -\frac{1}{b} \left(1 + \psi(1)\right) \\
-\frac{1}{b} \left(1 + \psi(1)\right) & \frac{a^2}{b^2},
\end{pmatrix};
\end{equation}
see \cite{gupta2006comparison}, but note the parametrization: some sources define the scale parameter as the reciprocal of what is given here.

The MLE properties of the Weibull distribution have been studied in
\cite{smith1985maximum, woodroofe1972maximum, akahira1975asymptotic}, although these works consider the three-parameter Weibull distribution for which MLE is especially problematic. It is shown that MLE is not even consistent if $a
\leq 1$. For $1 < a < 2$, MLE is not asymptotically normal, while
if $a=2$, MLE is asymptotically normal, but with different covariance
matrix than that of Theorem \ref{thm:an}. If $a > 2$, asymptotic
normality holds (as well as asymptotic efficiency). Note that the Weibull distribution is heavy-tailed (i.e. subexponential) if and only $a < 1$ (\cite{embrechts1997modelling}, Example 1.4.3 for the if statement, while reverse implication follows from the existence of a Cram\'er-Lundberg exponent when $a \geq 1$), hence MLE for subexponential Weibull distributions results in an inconsistent estimator.

The likelihood equations for the Weibull distribution can be solved explicitly. As will be seen in Section \ref{sec:MLEparam} (and \cite{mleorappendix}), for all four loss data sets to which we apply MLE, the ``true'' values of $\alpha$ are all less than one, leading to non-bounded Fisher information matrices. This manifests in the algorithm of \texttt{fitdistrplus} via warning messages that the resulting system is singular. In case of non-convergence of MLE, we discard the parameter estimate.

\subsubsection{Lognormal Distribution}
The lognormal distribution $\log \mathcal{N}(\mu, \sigma)$ has PDF
\begin{align*}
f(x|\mu, \sigma) &= \frac{1}{ \sqrt{2 \pi}\sigma x}\exp\left(-\frac{(\log x - \mu)^2}{2 \sigma^2} \right)
\end{align*}

Note that we choose the second parameter to be $\sigma$ and not $\sigma^2$. This convention makes a difference in calculating the Fisher information matrix, which is
\begin{equation}
I(\theta) = I(\mu, \sigma)=
\begin{pmatrix}
1/\sigma^2 & 0 \\
0 & 2/\sigma^2
\end{pmatrix}.
\end{equation}

The Fisher information matrix of the lognormal distribution is non-singular, since a random variable $X$ is lognormal if and only if there is a normally distributed random variable $Y$ with $X = \exp(Y)$, and the function $\exp: \RR \to (0, \infty)$ is a diffeomorphism (hence, loosely speaking, all properties involving derivatives and integrals that hold for normally distributed variables hold for lognormal, and vice-versa).

The lognormal distribution is also identifiable for all allowed $(\mu, \sigma)$, since the same holds for the normal distribution. The Fisher information matrix is positive-definite, since it is a diagonal matrix with positive entries. As with the Pareto distribution, the likelihood equations for the lognormal distribution can be solved explicitly, hence the determination of $(\widehat{\mu}, \widehat{\sigma})$ is computationally unproblematic.

\subsubsection{Log-logistic Distribution}
As with the lognormal distribution, a random variable $X$ follows a log-logistic distribution if and only if there exists a logistic random variable $Y$ such that $X = \exp(Y)$.
The log-logistic distribution $LL(a,s)$ has PDF
\begin{equation}
f(x | a,s) =\frac{ a \left(\frac{x}{s}\right)^a }{ x (1 + (x/s)^a)^2 }
\end{equation}
for $x\geq0$, and is zero otherwise. The parameters $a$ and $s$ must both be positive. Like the Pareto distribution, a log-logistic distributions can have an unbounded first moment, namely, when $a \leq 1$, while for $a \leq 2$, the variance is also unbounded.

The Fisher information matrix of $X \sim LL(a,s)$ is (\cite{llogFisher})
\begin{equation}
I(\theta) = I(a, s)=
\begin{pmatrix}
\frac{3+\pi^2}{9 a^2} & 0 \\
0 & \frac{1}{3} \left(\frac{a}{s}\right)^2
\end{pmatrix},
\end{equation}
which is positive-definite, since $a,s> 0$. Moreover, the log-logistic distribution is identifiable for $a > 1$, which follows from the above general strategy, since its median is $s$ and its mode is $s\left(\frac{a-1}{a+1} \right)^{1/a}$, which is strictly increasing in $a$, and is hence injective.

To fit the log-logistic distribution, we use \texttt{fitdistrplus}, which performs optimization with the Nelder-Mead algorithm. For initial parameter values, we take (see e.g. \cite{kthThesis})
\begin{align*}
s_{\textrm{init}} &=  \textrm{Median}(x_1, \ldots, x_n)\\
a_{\textrm{init}} &= \frac{\log(n-1)}{\log( \max(x_1, \ldots, x_n)/s_{\textrm{init}} })
\end{align*}

\subsubsection{Generalized Beta Distribution of the Second Kind}
The GB2 distribution (also known as the \emph{transformed Beta}
distribution) is nested within the more general Feller-Pareto 
distribution $FP(\mu,
\sigma, \gamma, \gamma_1, \gamma_2)$, and itself nests the Weibull, lognormal, and log-logistic
distributions (as well as the generalized Pareto and inverse Burr
distributions \cite{brazauskas2002fisher}), hence the GB2 distribution makes possible an evaluation of the trade-off between generality (GB2) and parsimony (Weibull, lognormal and log-logistic) when modeling OR loss data.

The GB2 distribution has PDF
\begin{equation}
f(x) = \frac{a(x/b)^{ap - 1}}{b B(p,q) (1 + (x/b)^a)^{p+q}},
\end{equation}
where $B(p,q)$ is the Beta function (or Euler integral), defined for $p,q > 0$ as
\begin{equation}
B(p,q) = \int_0^1 t^{p-1}(1-t){q-1} dt,
\end{equation}

The $m$th moment of of $X \sim GB2(a,b,p,q)$ is $\frac{b^m B(p+h/a, q-h/a)}{B(p,q)}$, and no moments are finite above the $aq$th one \cite{bookstaber1987general}. 

The Fisher information
matrix for the GB2 distribution can be derived from that of the Feller-Pareto distribution \cite{brazauskas2002fisher}. Specifically, since $\gbtwo(a,b,p,q) =
FP(0,b,1/a,q,p)$, we use the change-of-variable formula $J I_{FP}
J^{\top}$, where $J$ is the Jacobian matrix of the coordinate change
$\mu \to 0$, $\sigma \to b$, $\gamma \to 1/a$, $\gamma_1 \to q$, and
$\gamma_2 \to p$. Writing $I= (I_{i,j}) = (I_{j,i})$, the Fisher information
matrix for the $\gbtwo(a,b,p,q)$ distribution has entries
\begin{align*}
I_{1,1} &= a^2 + \frac{a^2 p q}{p+q+1}\\
I_{1,2} & = - \frac{pq\left(\psi(p) - \psi(q) \right) + q - p}{b(p+q-
  1)}\\
I_{1,3} & =\frac{q\left(\psi(p) - \psi(q) \right) - 1}{a(p+q)} \\
I_{1,4} & = \frac{p\left(\psi(q) - \psi(p) \right) - 1}{a(p+q)} \\
I_{2,2} & = \frac{a^2pq}{b^2(p + q + 1)} \\
I_{2,3} & = \frac{aq}{b(p+q)}\\
I_{2,4} & = \frac{-ap}{b(p+q)} \\
I_{3,3} & = \psi'(p) - \psi'(p+q) \\
I_{3,4} & = -\psi'(p+q) \\
I_{4,4} & = \psi'(q) - \psi'(p+q),
\end{align*}
where $\psi(x) = \Gamma'(x)/\Gamma(x)$ is the digamma function, and
its derivative $\psi'(x)$ is the trigamma function.

We implement our own MLE for the GB2 distribution as follows. We incorporate the linear bounds on $p,q$ by implicit penalty in the likelihood function, and minimize the negative log-likelihood function with the package \texttt{NelderMead} \cite{nm}. To obtain initial parameter values, we use the pseudo-MLE functionality of the package \texttt{GB2} \cite{gb2}. Since Nelder-Mead in general only returns a local minimum, we run the minimization at two other parameter start values, obtained by perturbing the loss data (losses shifted up for one, and shifted down for the other) and applying pseudo-MLE to the perturbed loss data. If at least one of the three start parameters leads to a convergent solution, we take the calibrated parameters corresponding to the lowest negative log-likelihood value.


Before turning to results, note that Theorem \ref{thm:an} assumes that we are able to find the global maximum of the likelihood function. For both the log-logistic and GB2 distributions, there is no guarantee of having found a global maximum. We hope that the descriptions of our methodology above will suffice to enable practitioners using the log-logistic or GB2 distributions to judge how their optimization algorithms differ.


 \section{Asymptotic normality for OpVaR severity distributions}
 \label{sec:MLEparam}

 In this section, we first evaluate asymptotic normality both graphically and numerically for the severity
 distributions described above when fitted to moderately heavy loss data in Sections \ref{sec:graphical} and \ref{sec:numerical}, respectively. In Section \ref{sec:anCIs}, we examine what implications this has for approximating parameter confidence intervals with asymptotic normality. The corresponding results for three other sets of loss data can be found in \cite{mleorappendix}.

To set the stage, we describe our simulation procedure (parametric boostrapping) in detail for a fixed loss data set \texttt{losses} and sample size $n$

For \texttt{distn} in $\{\texttt{pareto}, \texttt{Weibull}, \texttt{lognormal}, \texttt{log-logistic}, \texttt{GB2}\}$ with CDF $F(x|\theta)$
\begin{enumerate}
\item Fit \texttt{distn} to \texttt{losses} with MLE to obtain true parameters $\thtrue$
\item For $i$ in $\{1, \ldots, m\}$ (we take $m=40,000$)
\begin{enumerate}
\item Draw $n$ samples from the true distribution $F(x|\thtrue)$ to obtain bootstrapped losses $\texttt{losses}_i$
\item Fit \texttt{distn} to $\texttt{losses}_i$ with MLE to obtain bootstrapped parameters $\thhat_{i,n}$
\end{enumerate}
\end{enumerate}

For each distribution family and each sample-size $n$ we thus obtain statistics for parameter estimation. We then compare each component of the boostrapped parameters $\thhat_{1,n}, \ldots, \thhat_{m,n}$ to the prediction of Theorem \ref{thm:an}. For example, each $\thhat_{i,n}$ for a lognormal distribution will be a vector with two components, $\thhat_{i,n} = (\mu_{i,n}, \sigma_{i,n})$, and for each of these components we plot the corresponding kernel density estimate (essentially a smart histogram; see e.g. \cite{hastie2009elements}, Chapter 6) against the normal distribution of Theorem \ref{thm:an} for this component. Continuing with this example, the normal distribution corresponding to the bootstrapped $\mu_{i,n}$ will have variance $1/(\sigma^*)^2$. (For readibility, in the plots below we center the predicted normal distribution at the true value rather than 0, and move the factor of $\sqrt{n}$ to the right-hand side of the limit). Note that a comprehensive study of applicability of Theorem \ref{thm:an} would require more than the plots and confidence intervals we present, which only focus on the marginal distributions of the estimated parameters.


For the below example, the underlying loss data set consists of  losses, 19\% of which are above the splicing threshold of 100,000 EUR. The data are not particularly heavy, with mean of 131560, median of 39018, and no losses larger than 30m EUR. Results for heavier loss data are found in \cite{mleorappendix}.

\subsection{Graphical tests of asymptotic normality}
\label{sec:graphical}
\begin{figure}[b]
\begin{center}
\includegraphics{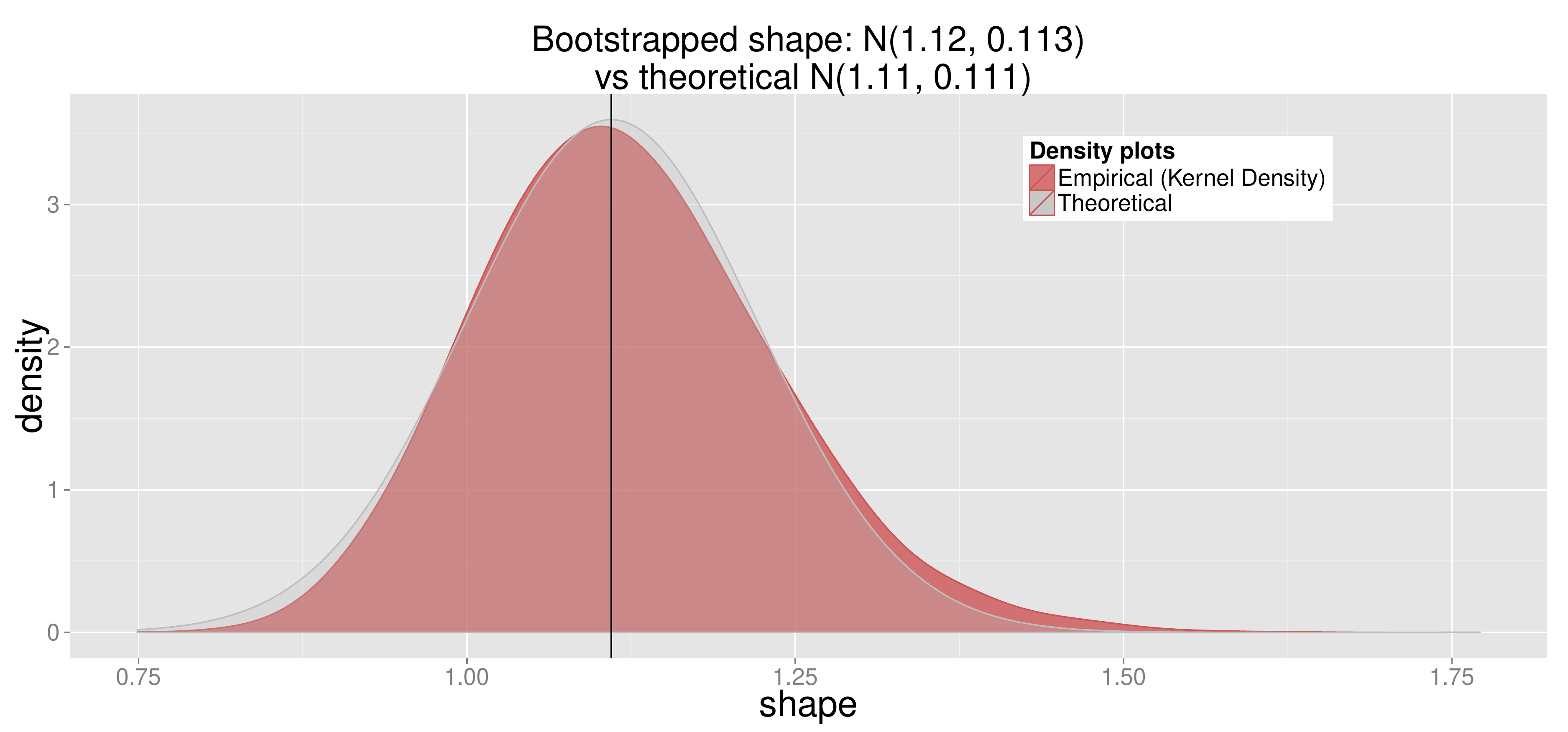}
\caption{\uomten: AN for Pareto: true $\theta =$ (1.11), sample-size 100 }
\label{fig:an10_pareto_100}
\end{center}
\end{figure}

\begin{figure}[b]
\begin{center}
\includegraphics{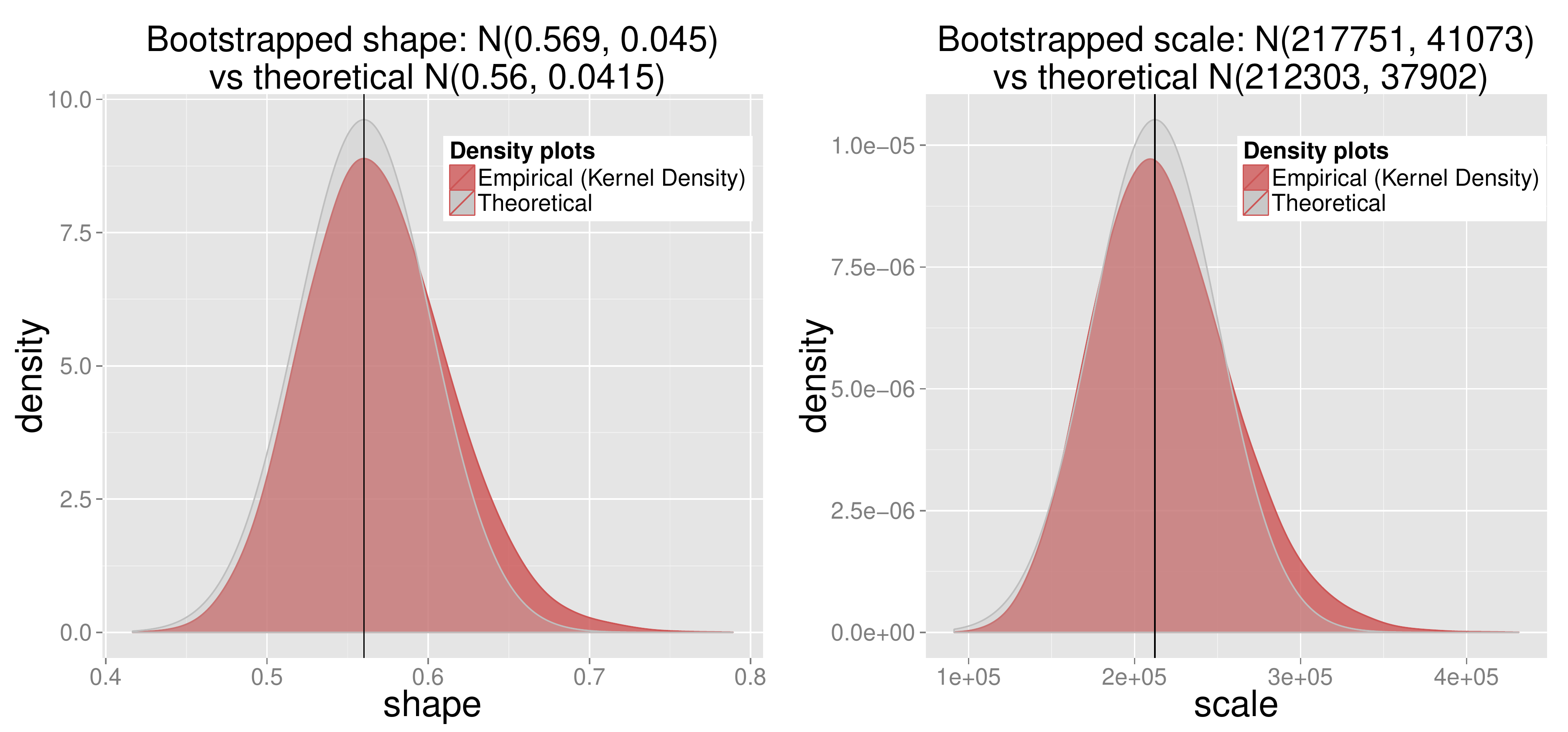}
\caption{\uomten: AN for Weibull: true $\theta =$ (0.56, 212303.18), sample-size 100 }
\label{fig:an10_weibull_100}
\end{center}
\end{figure}

\begin{figure}[b]
\begin{center}
\includegraphics{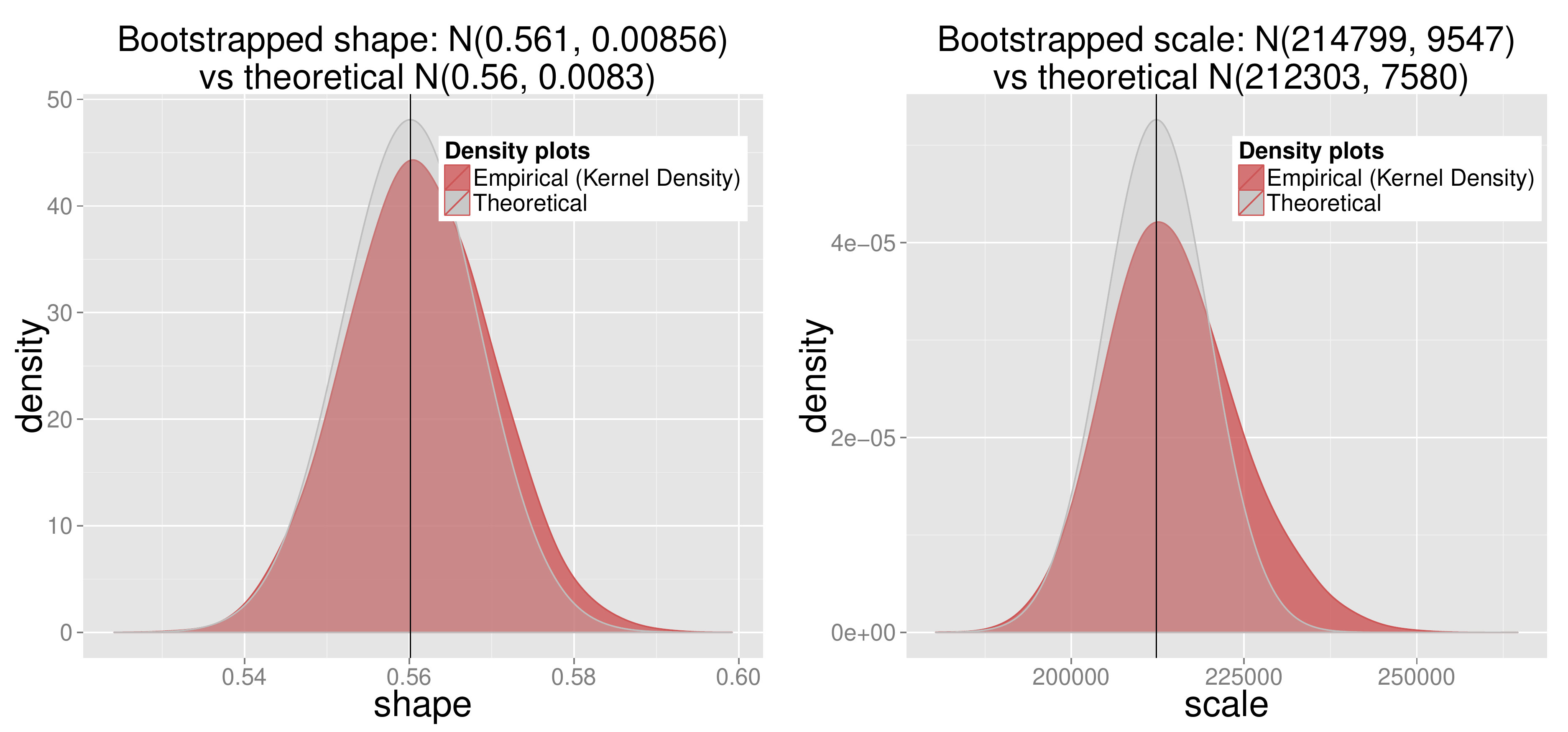}
\caption{\uomten: AN for Weibull: true $\theta =$ (0.56, 212303.18), sample-size 2500 }
\label{fig:an10_weibull_2500}
\end{center}
\end{figure}

\begin{figure}[b]
\begin{center}
\includegraphics{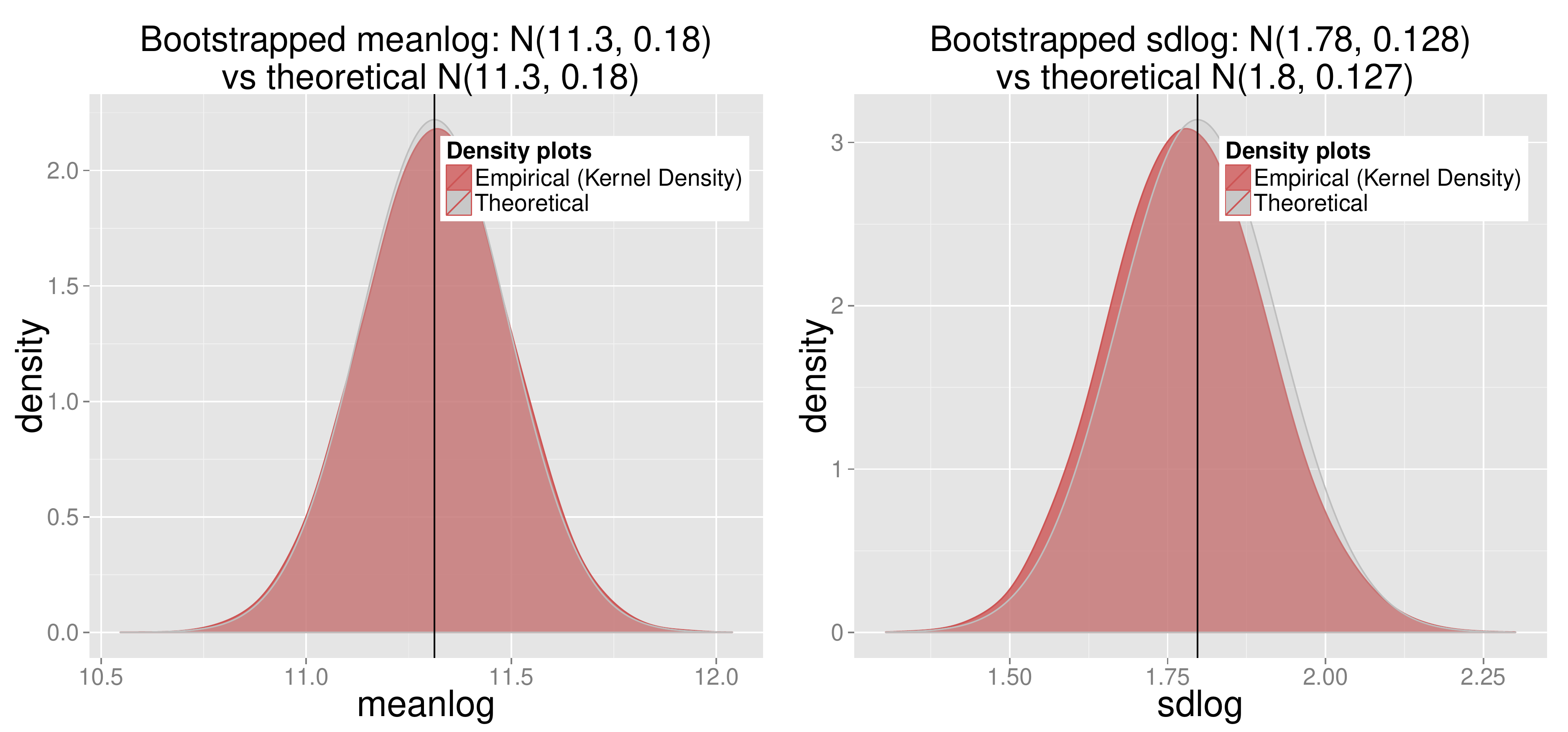}
\caption{\uomten: AN for lognormal: true $\theta =$ (11.3, 1.8), sample-size 100 }
\label{fig:an10_lnorm_100}
\end{center}
\end{figure}

\begin{figure}[b]
\begin{center}
\includegraphics{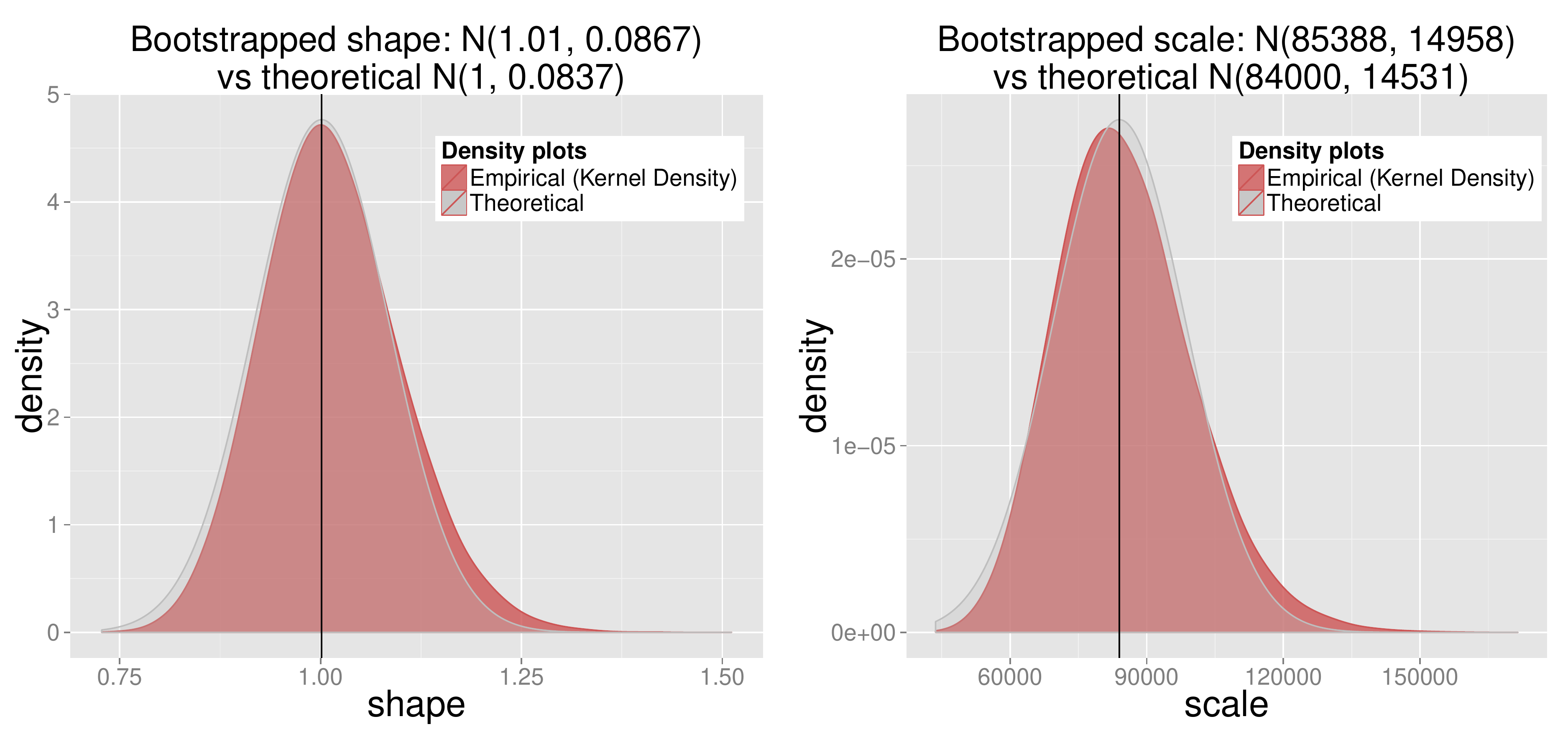}
\caption{\uomten: AN for log-logistic: true $\theta =$ (1, 84000), sample-size 100 }
\label{fig:an10_llogis_100}
\end{center}
\end{figure}

\begin{figure}[b]
\begin{center}
\includegraphics{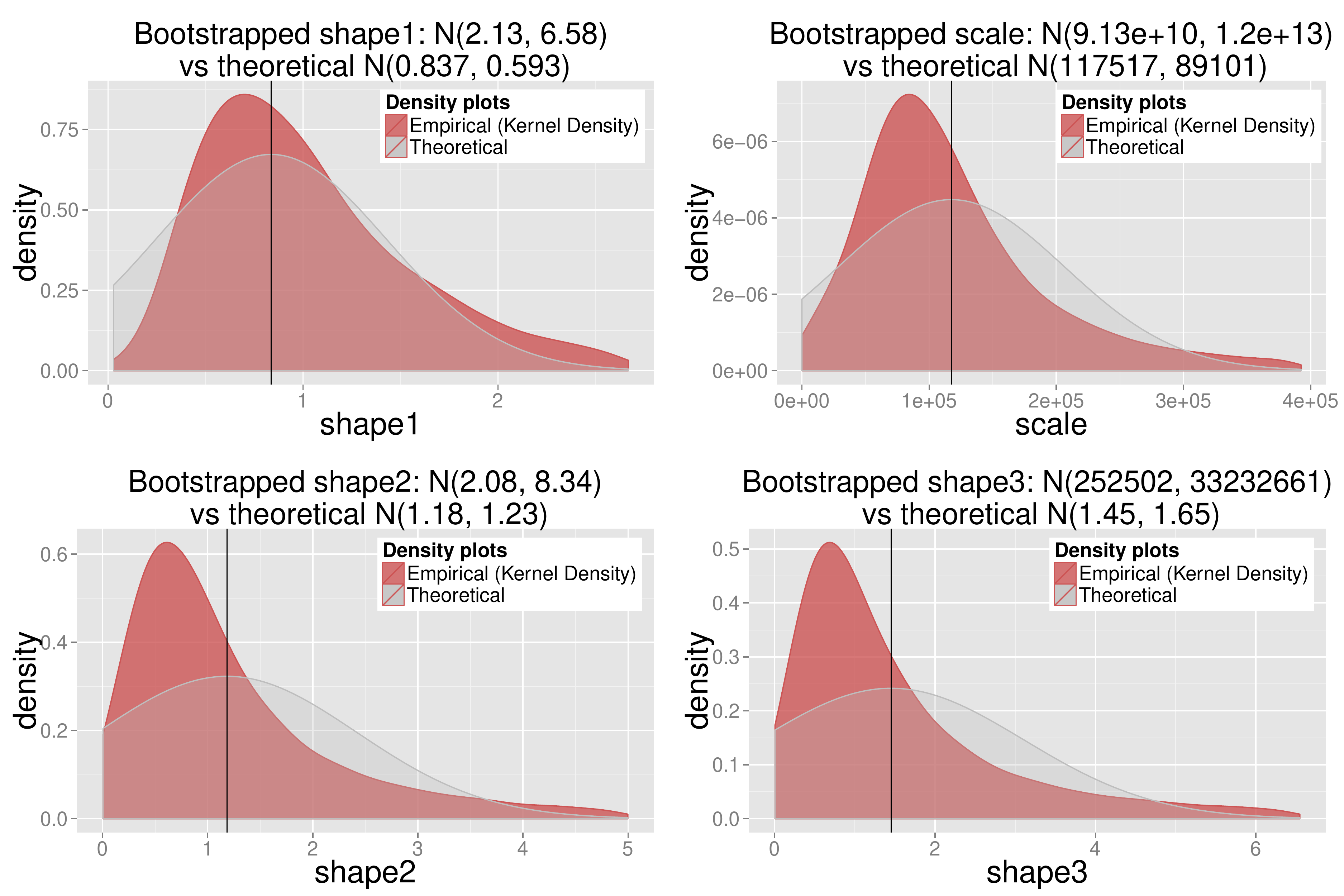}
\caption{\uomten: AN for GB2: true $\theta =$ (0.837, 117516.887, 1.184, 1.454), sample-size 100 }
\label{fig:an10_gb2_100}
\end{center}
\end{figure}

\begin{figure}[b]
\begin{center}
\includegraphics{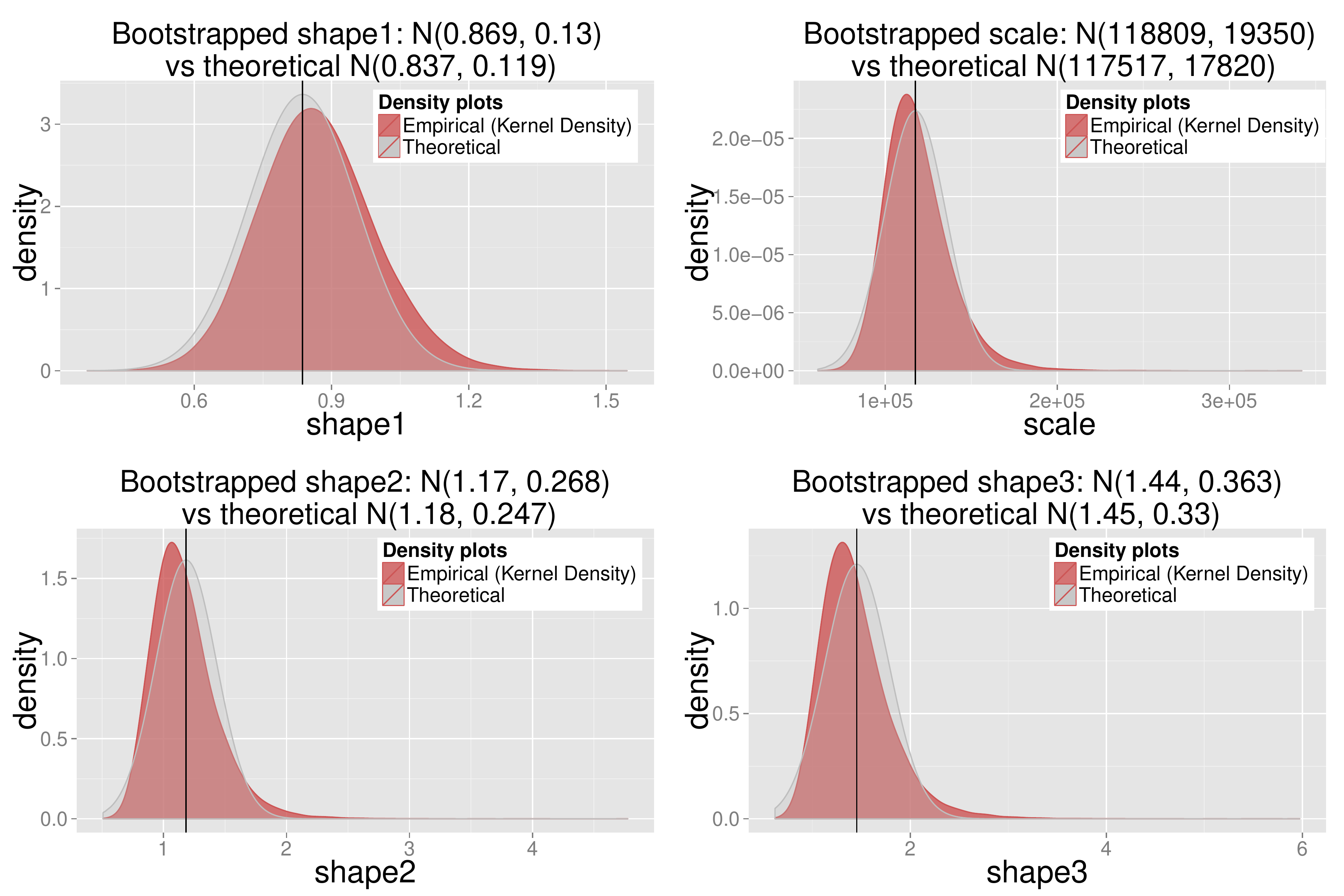}
\caption{\uomten: AN for GB2: true $\theta =$ (0.837, 117516.887, 1.184, 1.454), sample-size 2500 }
\label{fig:an10_gb2_2500}
\end{center}
\end{figure}

The graphical tests of normality of Figures \ref{fig:an10_pareto_100} -- \ref{fig:an10_gb2_2500} show how variablility in how normal the marginal parameter distribituions appear. On the `normal' side, the bootstrapped parameters behave as in Theorem \ref{thm:an} even for sample-size as small as $100$ for the Pareto (Figure \ref{fig:an10_pareto_100}) and lognormal distributions (Figure \ref{fig:an10_lnorm_100}. Nevertheless, for the Pareto distribution, the p-values tell a different story (see below), and for the lognormal distribution, the small-sample bias of the MLE estimator for $\sigma$ of the lognormal distribution can be seen).

Among the remaining distributions of Weibull, log-logistic and GB2, each shows varying degrees of skewness; see Figures \ref{fig:an10_weibull_2500}, \ref{fig:an10_llogis_100}, \ref{fig:an10_gb2_100}, \ref{fig:an10_gb2_2500}, respectively. The skewness for the Weibull distribution does not decrease with increasing sample-size, as it does for the others. This behavior is not surprising, however, since the true shape parameter is
$0.56$, i.e. the MLE is not consistent for this value, let alone asymptotically normal. The GB2 distribution for 100 tail losses shown in Figure \ref{fig:an10_gb2_100} gives by far the worst match to asymptotic normality.





\subsection{Numerical tests of asymptotic normality}
\label{sec:numerical}

The graphical tests above consider only the marginal boostrapped parameter distributions (i.e. is each MLE-fitted parameter indidually normally distributed as predicted by theory?). To fully test Theorem \ref{thm:an}, we turn to numerical tests of normality.

Hypothesis testing for normally distributed data has its merits and demerits as a modeling tool, but it is nevertheless a feature of the regulatory statistical landscape. The current task of assessing the validity of Theorem \ref{thm:an} based on simulated data is quite different from the usual situation of having real-world data. In the real-world, it is safe to assume that no large data set is truly normally distributed, thus diminishing the value of normality tests for large samples sizes. In the present context, we also know that our 40000 bootstrapped parameters are not normally distributed, but if Theorem \ref{thm:an} holds, then increasing sample sizes (i.e. assumed size of loss data) will lead to fewer rejections of the normality hypothesis, or, equivalently, bigger p-values.

For the Pareto distribution, we use the Anderson-Darling test as implemented in \texttt{nortest} \cite{nortest}. This test has the advantage of working for all sample-sizes, unlike the Shapiro-Wilk test. All p-values except for 2500 samples are indistinguishable from 0 (i.e. smaller than $10^{-12}$). For a sample-size of 2500, the p-value is $2.28e-09$.

For the remaining multivariable distributions, we use the Mardia test \cite{mardia1} as implemented in \texttt{MVN} \cite{Korkmaz:2014fk}. Unlike the Anderson-Darling test, The Mardia test checks skewness and kurtosis separetely. The only distribution with skew p-values greater than $10^{-15}$ is the lognormal distribution. Its p-values are shown in table \ref{tab:skew}, where we see that the typical $5\%$ significance level would not reject the normal hypothesis for samples sizes 1500 and 2500.

The kurtosis test leads to essentially 0 p-values with the GB2 distribution. The kurtosis p-values for the other distributions are shown in Table \ref{tab:kurt}. A $5\%$ significance level would fail to reject the normal hypothesis for the lognormal distribution across the whole range of sample-sizes considered here, and likewise for the log-logistic distribution for the 1000 and 1500 sample-sizes.

The Weibull distribution is the one that most clearly fails the assumptions of Theorem \ref{thm:an}, so it is no surprise that the kurtosis p-values do not increase monotonically. The expected monotone increase of kurtosis p-values with increasing sample sizes is not apparent for the lognormal distribution. This behavior could result from numerical instability for such large data sets.


\begin{table}[ht]
\centering
\caption{Mardia skew p-values per sample-size} 
\label{tab:skew}
\begin{tabular}{rlrrrrrr}
  \hline
 & 100 & 200 & 300 & 500 & 1000 & 1500 & 2500 \\ 
  \hline
lognormal & 0 & 0.01 & 0.03 & 0.03 & 0.17 & 0.32 & 0.26 \\ 
   \hline
\end{tabular}
\end{table}
\begin{table}[ht]
\centering
\caption{Mardia kurtosis p-values per sample-size} 
\label{tab:kurt}
\begin{tabular}{rrrrrrrr}
  \hline
 & 100 & 200 & 300 & 500 & 1000 & 1500 & 2500 \\ 
  \hline
Weibull & 0.00 & 0.00 & 0.00 & 0.00 & 0.02 & 0.00 & 0.00 \\ 
  lognormal & 0.70 & 0.44 & 0.88 & 0.36 & 0.81 & 0.70 & 0.57 \\ 
  log-logistic & 0.00 & 0.00 & 0.00 & 0.00 & 0.55 & 0.36 & 0.00 \\ 
   \hline
\end{tabular}
\end{table}
In contrast to the p-values presented in this section, the next section looks beyond whether or not Theorem \ref{thm:an} holds to study the validity of the commonly-used normal approximation to estimate parameter confidence intervals.

\subsection{Approximating parameter confidence intervals}
\label{sec:anCIs}
We now turn to question \ref{enum:conf} from the introduction about the use of Theorem \ref{thm:an} to approximate parameter confidence intervals. Several methods exist for evaluating the goodness-of-fit of a severity distribution to loss data. The one we focus on here follows directly from asymptotic normality. Assuming that Theorem \ref{thm:an} holds, then the MLE fitted parameters are normally distributed about the ``true'' (i.e. fitted) parameters, with covariance matrix determined by the Fisher information matrix (see Section \ref{sec:an}). As we have seen already, however, this normal assumption is questionable for small-sample sizes of operational risk data. In this section, we quantify the resulting confidence interval error.

The below table shows the percent error of using this approximation relative to the ``true'' 95\% confidence intervals obtained by quantiles of the boostrapped parameters, i.e. we compare the difference of the 97.5\% quantile and 2.5\% quantile from 40,000 bootstrapped parameters to same difference of quantiles from the normal distribution dictated by Theorem \ref{thm:an}.

The results in Table \ref{tab:confInts} mirror what can be seen from the plots: for the Pareto, lognormal and log-logistic distributions, the normally approximated 95\% confidence intervals are within a few percent of the true ones, while the approximation is relatively poor for the Weibull and GB2 distributions. For GB2, the normally approximated confidence intervals are within 10\% of the true ones given enough data ($2500$ losses). The approximation for the Weibull distribution gets worse as sample sizes increase, a phenomenon that can also be seen from Figures \ref{fig:an10_weibull_100} and \ref{fig:an10_weibull_2500}.

\begin{table}[ht]
\centering
\caption{Percent error of 95\% confidence intervals derived from asymptotic normality by sample-size} 
\label{tab:confInts}
\begin{tabular}{rlllllll}
  \hline
 & 100 & 200 & 300 & 500 & 1000 & 1500 & 2500 \\ 
  \hline
Pareto shape & 2\% & 1\% & 0\% & 0\% & 1\% & 0\% & -1\% \\ 
  Weibull shape & 8\% & 7\% & 7\% & 6\% &  6\% &  5\% &  6\% \\ 
  Weibull scale & 9\% & 8\% & 9\% & 10\% & 13\% & 16\% & 22\% \\ 
  lognormal meanlog & 0\% & 1\% & 0\% & 1\% & 0\% &  0\% & 0\% \\ 
  lognormal sdlog & 0\% & 0\% & 0\% & 0\% & 0\% & -1\% & 0\% \\ 
  log-logistic shape & 3\% & 1\% & 0\% & 0\% & 0\% & 0\% & -1\% \\ 
  log-logistic scale & 2\% & 2\% & 1\% & 1\% & 0\% & 1\% &  0\% \\ 
  GB2 shape1 & 73\% & 31\% & 21\% & 15\% & 11\% &  8\% & 7\% \\ 
  GB2 scale & 79\% & 78\% & 73\% & 61\% & 31\% & 17\% & 7\% \\ 
  GB2 shape2 & 43\% & 50\% & 50\% & 45\% & 26\% & 15\% & 6\% \\ 
  GB2 shape3 & 43\% & 55\% & 58\% & 53\% & 30\% & 17\% & 7\% \\ 
   \hline
\end{tabular}
\end{table}

\clearpage
\section{Conclusions}
\label{sec:cons}
The only severity distribution considered here for which asymptotic normality clearly holds--even for sample sizes typical of operational loss data--is the lognormal distribution. Recall that MLE for it, along with the Pareto distribution, can be performed analytically (i.e. solving a simple equation). More generally, all key assumptions of Theorem \ref{thm:an} can be effectively verified for the lognormal distribution. Asymptotic normality also gives very good approximations for MLE parameter confidence intervals, with errors no larger than $1\%$ across all sample sizes.

The Pareto and log-logistic distributions fare similarly well, with normally approximated confidence intervals leading to errors no more than $3\%$. In contrast to the lognormal distribution, however, the Anderson-Darling p-values for the  Pareto shape parameter indicated a rejection of the normal hypothesis for all sample sizes. 

The generalized Beta distribution of the second kind (GB2) fares the worst in all tests of asymptotic normality. When it comes to desirable MLE properties and confidence intervals, this distribution should only be used with great caution.

A more interesting ``under-performer'' here is the Weibull distribution. In Section \ref{sec:basics}, we show that a Weibull distribution is sub-exponential precisely when its MLE is inconsistent. The graphical and numerical tests of asymptotic normality bear out this problem. The normal approximation error for MLE confidence intervals even gets worse as the sample size increases for the scale parameter. Although these analyses give clear warning signals about the MLE properties of Weibull as an operational risk severity distribution, we will show in \cite{mleor_var} that its stability properties as a function of sample size are good when compared to the other distributions.

The generally positive results for using asymptotic normality to approximate parameter confidence intervals should be taken also in the broader context operational risk modeling, with particular detail to the assumptions our simulation study makes and how this relates to real-world loss data.

Parametric bootstrapping is by definition compatible with one of the key assumptions of MLE, namely that data are independent and identically distributed, since the bootstrapped data samples are drawn independently from a single ``true'' distribution. That these assumptions hold for actual loss data has been questioned, and the resulting impact for loss data that are not independent and identically distributed on MLE is a focus of \cite{opdyke2012estimating}. Operational risk literature is not unanimous on this question, but the regular assessments of the ORX loss data consortium offer general backing for this assumption \cite{cope2008observed, orx2}. Besides the assumption of large sample-sizes, we investigate assumptions on distribution being fitted. It is, however, well-known that MLE can exhibit asymptotic normality even when the assumptions of the theory are not met (see e.g. \cite{smith1985maximum}). Loosely put, the assumptions are necessary to prove the result, but not necessarily for the result to hold. For three of the distributions (Pareto, lognormal, and log-logistic), we see high levels of agreement between theory and simulation, even for very small sample-sizes. The assumptions of asymptotic normality for the generalized Beta distribution of the second kind are difficult to verify.

We have also assumed implicitly the standard MLE requirement that we know the ``true'' underlying distribution. Robustness of a fitting method to misspecified models (i.e. selecting the wrong ``true'' distribution) is itself a topic of research; see e.g. \cite{ergashev2008should, opdyke2012estimating}.


\section{Declarations of interest and acknowledgments}
The opinions expressed in this article are those of the author and do not reflect the views of Allianz SE or Deutsche Bank AG. The author alone is responsible for the content and writing of the paper.

This research was conducted while I was employed by Deutsche Bank. It is a pleasure to thank Michael Kalkbrener, both for his support of this project and his input throughout. I would like to express my gratitude to Fabio Piacenza, who provided me with many insights and helpful suggestions for improvements. I would also like to thank J\"org Fritscher, Haider Haider, Tsz-Yan Lam, and Grygoriy Tymchenko for useful conversations and suggestions. I gratefully acknowledge permission to use loss data from the Operational Riskdata eXchange Association (\href{https://www.orx.org/Pages/HomePage.aspx}{ORX}), and would like to thank Luke Carrivick of ORX for his support and feedback.

\bibliography{../db_pll}
\bibliographystyle{apalike}

\end{document}